\documentclass[a4paper]{article}

\usepackage{INTERSPEECH2021}
\usepackage{url}
\usepackage{multirow}  
\usepackage{subfig}
\newcommand{\at}{\makeatletter @\makeatother}

\title{IR-GAN: Room impulse response generator for far-field speech recognition
}
\name{Anton Ratnarajah$^1$, Zhenyu Tang$^1$, Dinesh Manocha$^1$}
\address{
  $^1$University of Maryland, College Park, MD 20742, United States}
\email{jeran@umd.edu,zhy@umd.edu,dmanocha@umd.edu}

\begin{document}

\maketitle
\begin{abstract}
We present a Generative Adversarial Network (GAN) based room impulse response generator (IR-GAN) for generating realistic synthetic room impulse responses (RIRs). IR-GAN  extracts acoustic parameters from captured real-world RIRs and uses these parameters to generate new synthetic RIRs. We use these generated synthetic RIRs to improve far-field automatic speech recognition in new environments that are different from the ones used in training datasets. In particular, we augment the far-field speech training set by convolving our synthesized RIRs with a clean LibriSpeech dataset \cite{LibriSpeech}. We evaluate the quality of our synthetic RIRs on the real-world LibriSpeech test set created using real-world RIRs from the BUT ReverbDB \cite{ButReverb} and AIR \cite{AIR} datasets. Our IR-GAN reports up to an 8.95\% lower error rate than Geometric Acoustic Simulator (GAS) in far-field speech recognition benchmarks. We further improve the performance when we combine our synthetic RIRs with synthetic impulse responses generated using GAS. This combination can reduce the word error rate by up to 14.3\% in far-field speech recognition benchmarks.
\end{abstract}

\section{Introduction}
\label{sec:intro}

Reverberation is a part of the speech signal which characterizes the acoustic environment (e.g., room geometry, loudspeaker and microphone location, room materials, etc.) used to capture the speech signal. Reverberation can be characterized by the transfer function known as the room impulse response (RIR). A room impulse response represents the relationship between the dry sound and the reflection of the sound signal from the boundaries of the room \cite{room_acoustics_vigran_2014}. 


RIRs are frequently used in many practical applications such as far-field speech recognition~\cite{ButReverb,low-frequency_zhenyu,nspeech1}, speech enhancement~\cite{speech_enhan2}, speech separation~\cite{cone-of-silence}, sound rendering~\cite{sound_rendering}, audio forensics~\cite{audio1}, etc. One challenge for these applications is that existing recorded or real-world RIR datasets are collected in limited acoustic environments. 
In our paper, we address this issue by augmenting RIRs covering a wide range of acoustic environments using a Generative Adversarial Network (GAN). 


\begin{figure}[t] 
	\centering
	\includegraphics[width=2.5in]{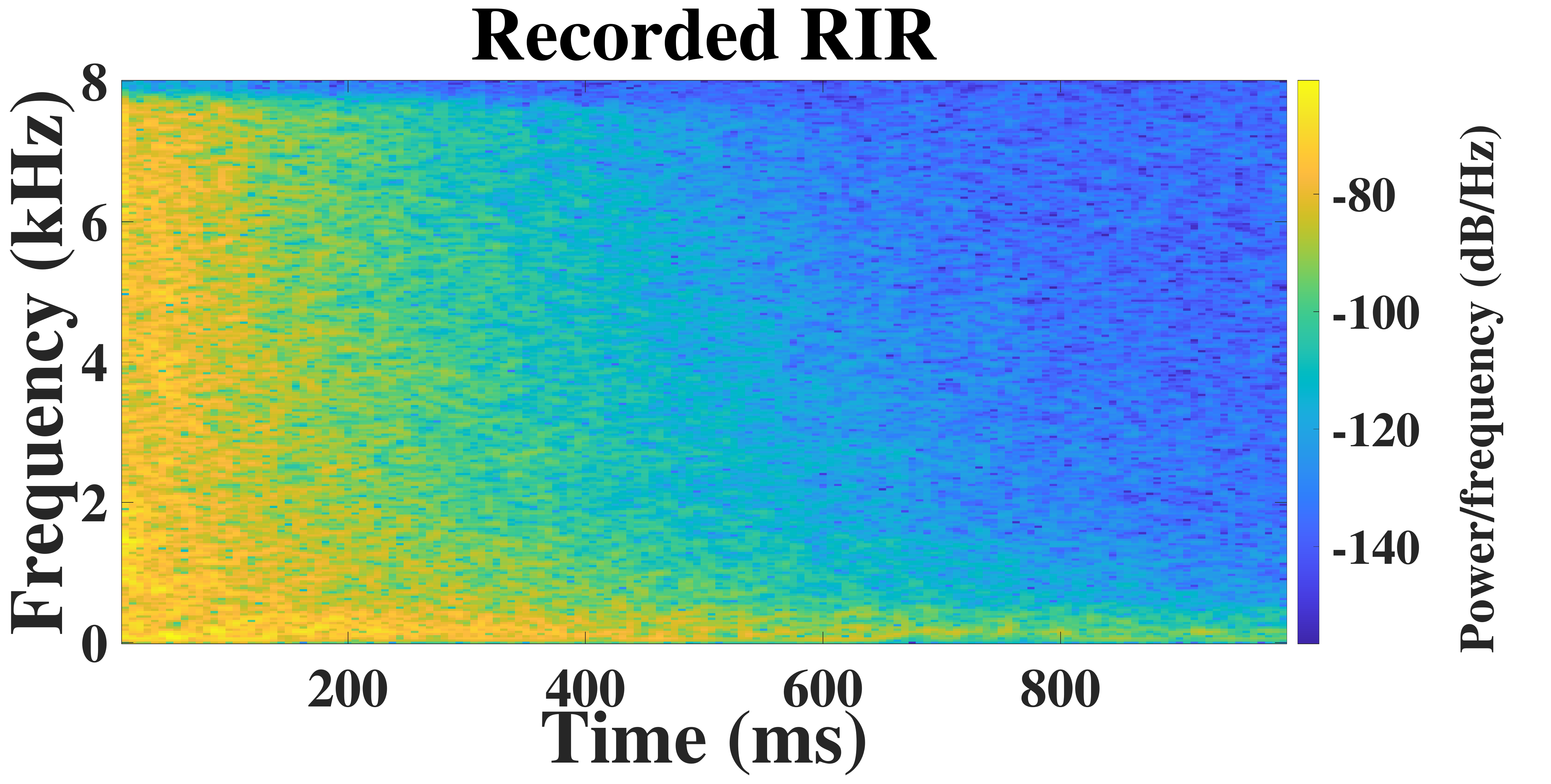}
	\includegraphics[width=2.5in]{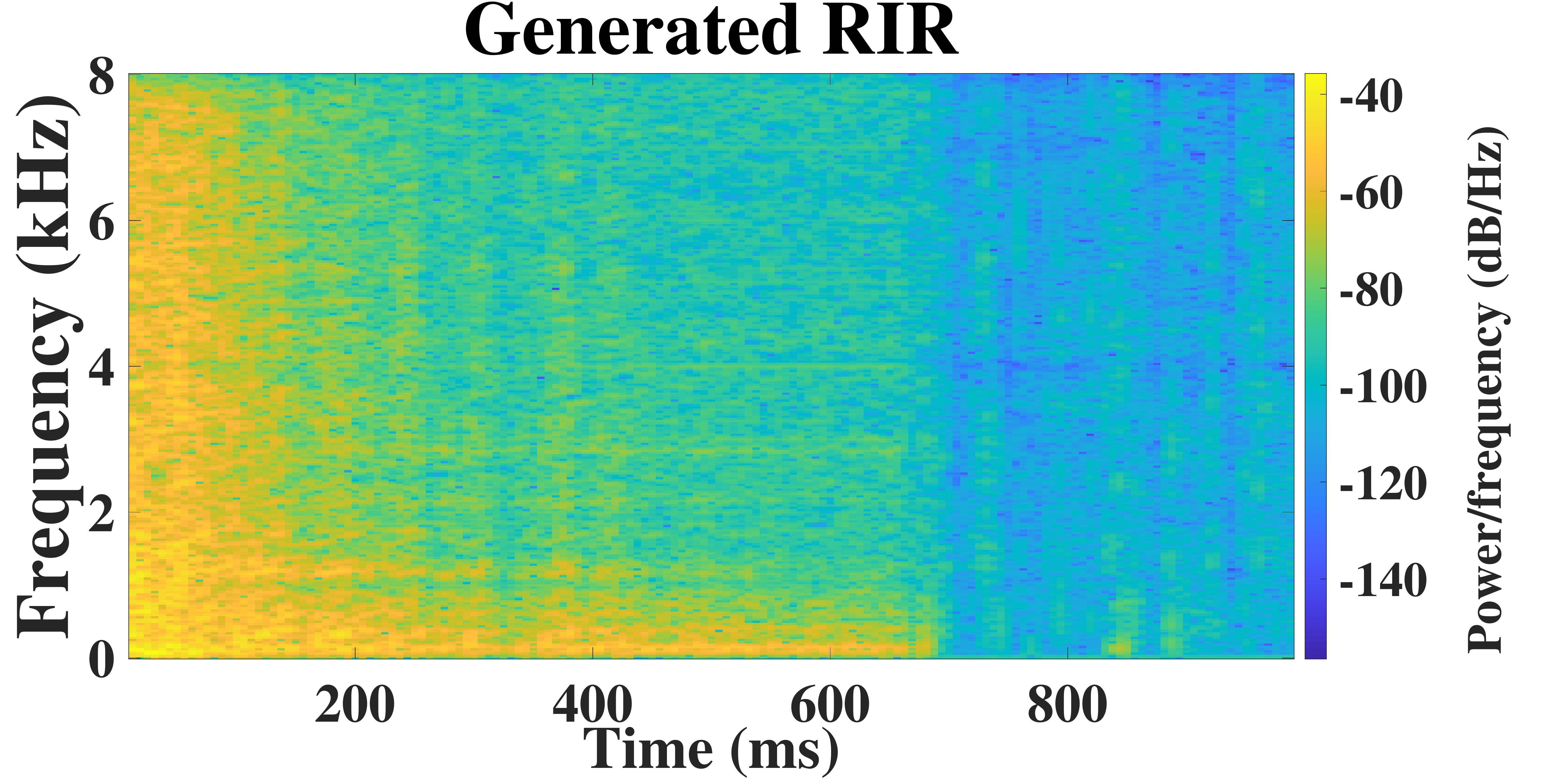}	
	\caption{Spectrogram of real RIR and RIR generated using our GAN-based approach. We can see both spectrograms have similar energy distributions.}
	\label{fig_1}
\end{figure}
Many prior techniques for generating synthetic RIRs are based on acoustic simulators~\cite{allen1979image,Diffuse_Acoustic_Simulation}. These simulators use room geometry, sound absorption, and sound reflection coefficients as input and generate RIRs by simulating occlusion, specular reflections and diffuse reflections. In practice,
such synthetic RIR generators can model some sound propagation phenomena in regularly shaped or mostly empty rooms. On the other hand, simulating the sound reverberation effects in complex scenarios like stairways is more difficult~\cite{low-frequency_zhenyu}. As a result, we need other synthetic RIR generator methods that can model the sound effects in complex environments.

{\bf Main Contributions:}  
We present a novel GAN-based RIR generator (IR-GAN) that is trained on real-world RIRs.
IR-GAN can parametrically control different acoustic parameters (e.g., reverberation time, direct-to-reverberant ratio, etc.) learned from real RIRs and generate synthetic RIRs that can imitate new or different acoustic environments. Moreover, we propose a constrained RIR generation approach that can avoid synthesizing RIRs with noisy artifacts to a greater extent.

Our IR-GAN uses WaveGAN \cite{wavegan} architecture to synthesize new RIRs by learning from real RIRs. IR-GAN maps all acoustic parameters in real RIRs to a high-dimensional space and generates a wide range of RIRs by controlling various acoustic parameters. As a result, we can train on real-world RIRs corresponding to complex locations like stairways and can augment RIRs corresponding to such locations. Figure \ref{fig_1} highlights the spectrogram of a real RIR from the BUT ReverbDB dataset \cite{ButReverb} and the spectrogram of a generated RIR using our approach.

We create far-field speech to evaluate our RIRs in a far-field automatic speech recognition (ASR) system. Our far-field speech $x_{f}[t]$ is augmented by convolving clean speech $x[t]$ from the LibriSpeech dataset \cite{LibriSpeech} with RIRs $h[t]$ and adding background noise $n[t]$ from the BUT ReverbDB \cite{ButReverb} dataset. The starting position $k$ of the noise is selected randomly, and the noise is repeated in a loop to fill clean speech. The weight $\alpha$ is calculated for a random signal-to-noise ratio within the range 10 to 100 for each far-field speech. We use real RIRs from the BUT ReverbDB \cite{ButReverb} and AIR \cite{AIR} datasets to create real-world far-field speech. 

{
\begin{equation}\label{eq:speech}
\begin{aligned}[b]
x_{f}[t] =  x[t] \circledast h[t] + \alpha * n[t + k].  
\end{aligned}
\end{equation}
}


Our GAN-based synthetic RIR generation approach is complementary to prior synthetic RIR generators based on acoustic simulators.  We evaluate the performance by conducting far-field ASR tests and show that combining RIRs synthesized from IR-GAN and a state-of-the-art geometric acoustic simulator reduces the word error rate by up to 14.3\%. Our code is published for follow-up research~\footnote{\url{https://gamma.umd.edu/pro/speech/ir-gan} (with video)}.

\section{Related Works}
\label{sec:method}



Physically-based acoustic simulators have been used over the decades to generate synthetic RIRs for far-field speech research. Wave-based methods and geometric methods are widely used to model RIRs for different acoustic environments. The wave-based approach \cite{wave1} solves the wave equation using numerical methods. Although wave-based methods accurately compute the RIRs, these methods are computationally expensive and only feasible for low frequencies and less complex scenes. The image source method \cite{image_method} and path tracing methods \cite{Diffuse_Acoustic_Simulation,geometric1} are common geometric acoustic simulation-based methods. The image source method only models specular reflections in simple rectangular rooms while path tracing-based geometric acoustic simulators model occlusion and specular and diffuse reflections. Geometric acoustic simulators treat sound waves in the form of a ray \cite{ray_assume1}. Although this assumption holds for high-frequency waves, ray assumption causes visible irregularities at low frequencies \cite{low-frequency_zhenyu}. In recent works, TS-RIRGAN \cite{anton-cycle} is used to transfer low-frequency wave effects learned from real RIRs to synthetic RIRs generated using geometric acoustic simulators. In many scenarios, the target room environment (i.e., the exact geometric shape and material parameters) is unknown or too complex for the geometric acoustic simulators. As a result, their ability to generate RIRs for all kinds of scenarios can be limited.

To overcome the limitations of synthetic RIRs, real RIRs are recorded in a controlled environment. The maximum length sequence method \cite{Maximum-Length-Sequence}, the time-stretched pulses method \cite{Time-Stretched-Pulses}, and the exponential sine sweep method \cite{Sine_Sweep} are common methods to measure real RIRs. Among these approaches, the exponential sine sweep method is robust to changing loudspeaker output volume and performs well in automatic speech recognition tasks. The real RIRs in BUT ReverbDB \cite{ButReverb} are collected using the exponential sine sweep method. Since collecting real RIRs is time-consuming and technically difficult, only a limited number of real RIRs is available to augment far-field speech.  

GANs have made steady progress over the years in image generation \cite{image1_nvidia}, image inpainting \cite{impainting}, and domain adaptation \cite{cyclegan}. The success of GAN in computer vision motivated researchers to use it in other fields. Recently, GANs have been becoming popular in the audio generation. GANs have shown progress from music generation \cite{MusicGAN} to any short audio clip generation \cite{wavegan, MelGAN}. In this work, we aim to augment high-quality RIRs using existing real RIRs. We use GANs for RIR generation to complement the prior works.

\section{OUR APPROACH: IR-GAN}
\label{sec:propose_method}
\subsection{Room Impulse Response Statistics}
\label{ssec:statistics}
Room impulse response acoustic parameters are used to characterize the acoustic environment \cite{bryan_acoustic_parameter} and control RIR generation using GAN. Reverberation time ($T_{60}$), direct-to-reverberant ratio (DRR), early-decay-time (EDT), and early-to-late index (CTE) are four acoustic parameters that can be estimated from RIRs. We use these acoustic parameters to constrain IR-GAN-based RIR augmentation. Reverberation time measures the amount of time taken to decay the sound pressure by 60 decibels (dB). The $T_{60}$ value depends on room size and the characteristics of the material (e.g., floor, walls, furniture, etc.). DRR is calculated by dividing the sound pressure level of a direct sound source by the sound pressure level of the sound arriving after one or more surface reflections \cite{drr_book}. DRR is measured in dB. Time taken for sound pressure to decay by 10 dB is multiplied by a factor of 6 to get early-decay-time. EDT depends on the type and location of the sound source. CTE measures the proportion of the total sound energy received in the first $50 ms$ to the energy received during the rest of the period \cite{room_acoustics_vigran_2014}.

\subsection{Room Impulse Response Representation}
\label{ssec:RIR_represent}
The representation of the input data and the data generated by the neural network is important for synthesizing high-quality RIRs. Therefore, lossy representations of RIRs like Mel-frequency cepstral coefficients (MFCCs) are less favorable. Audio samples are a lossless representation that can be easily converted to an audio signal. As different datasets store RIRs with different sampling rates, we re-sample all the RIRs to 16 kHz. We pass audio samples as a 32-bit floating-point vector of length 16384 to the GAN. The vector length is sufficient to represent RIRs because most of the real RIRs are less than one second in duration. We can represent slightly more than one second with 16384 samples with a sampling rate of 16 kHz.
\begin{figure}[t] 
	\centering
	\includegraphics[width=2.5in]{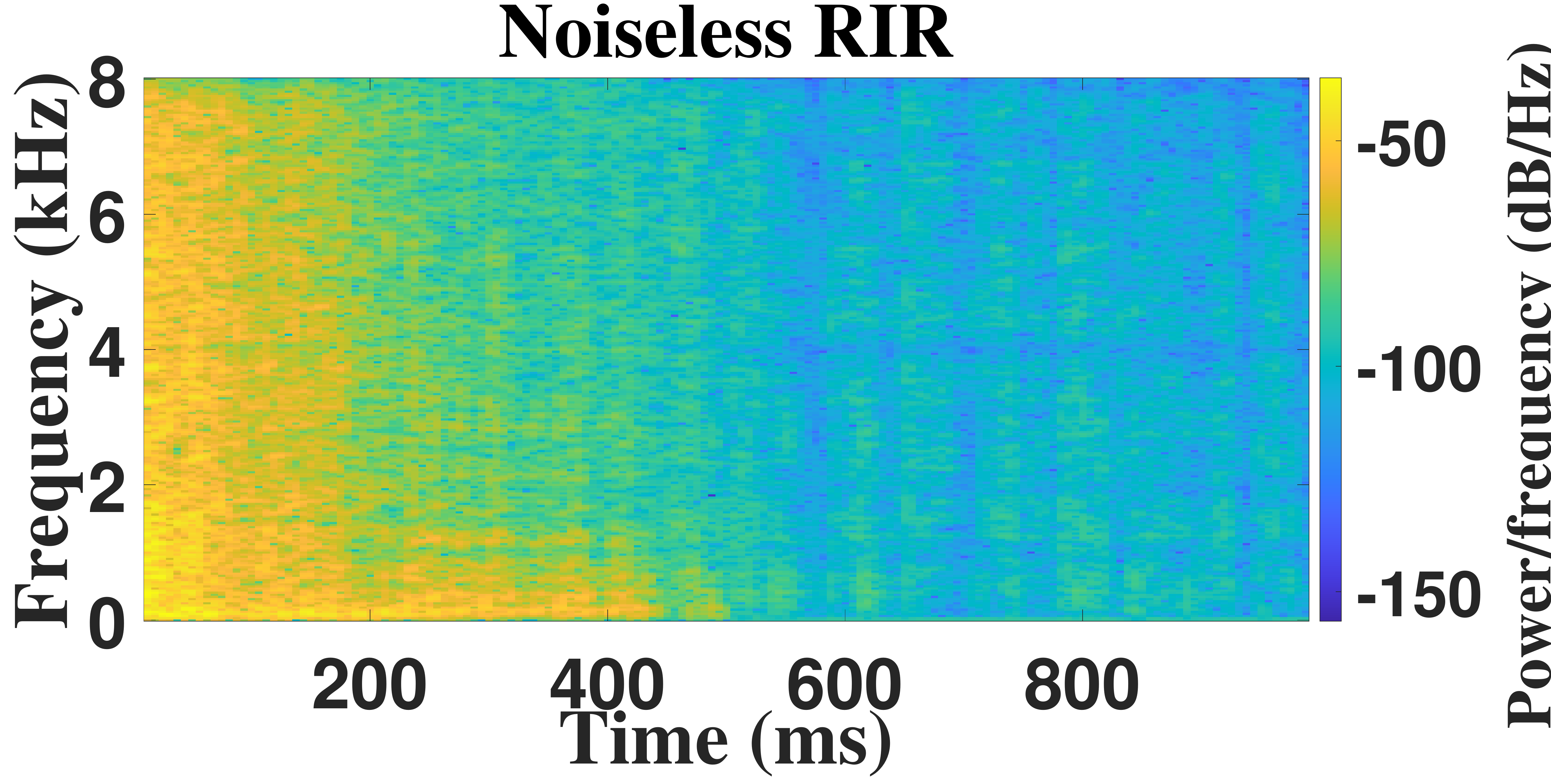}
	\includegraphics[width=2.5in]{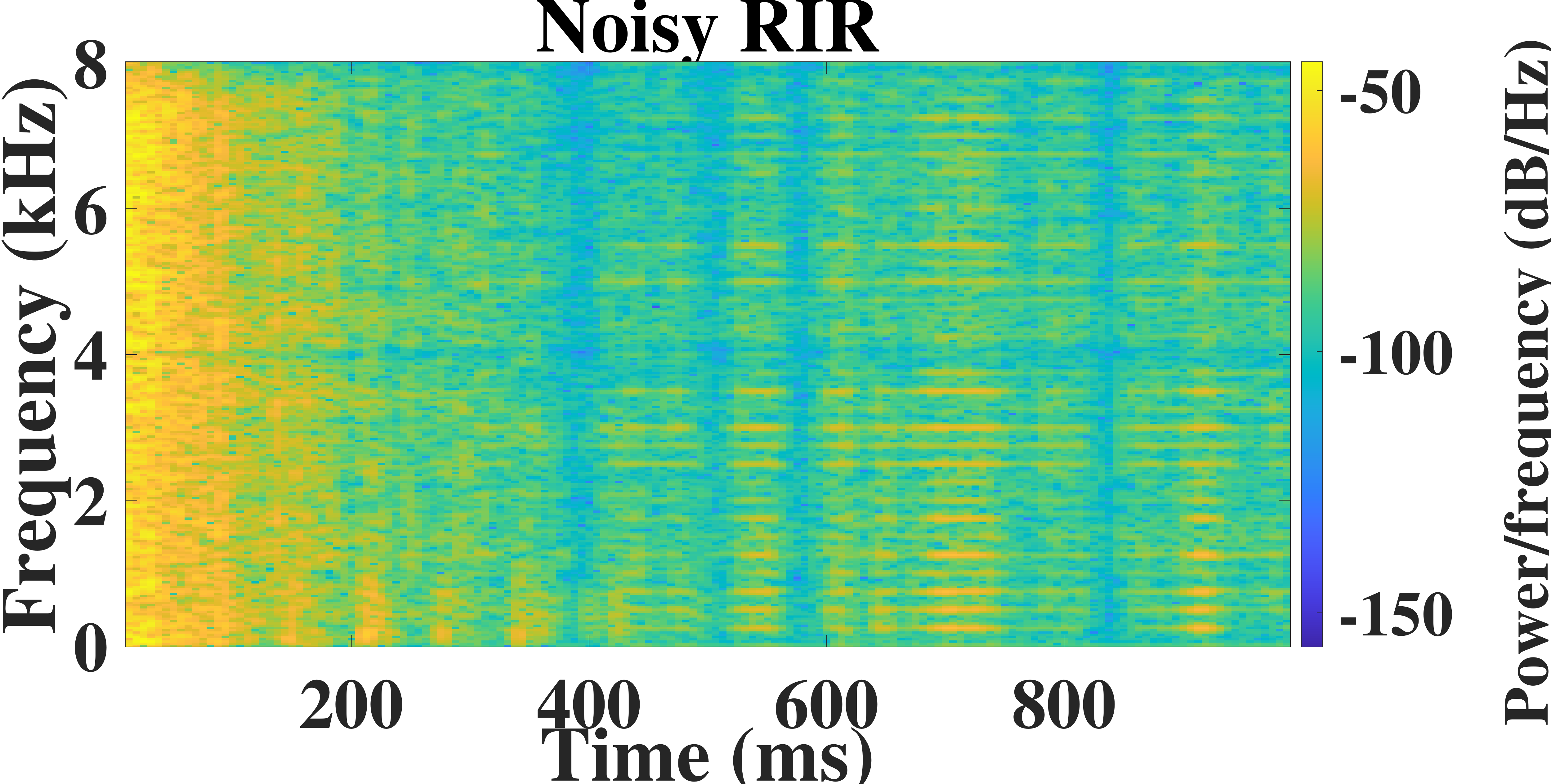}	
	\caption{Spectrogram of noiseless RIR and noisy RIR. The noiseless RIR has a $T_{60}$ value of around 1, and the noisy RIR has a $T_{60}$ value of around 3. In the noisy spectrogram, we can see many horizontal artifacts around 700ms.}
	\label{fig_2}
\end{figure}

\subsection{GAN}
\label{ssec:GAN}
GAN is a generative model that learns a mapping from a low-dimensional vector space to a high-dimensional space where the data is represented. We adapt the WaveGAN architecture proposed in \cite{wavegan} to generate high-quality RIRs. WaveGAN is a one-dimensional version of DCGAN \cite{DCGAN} where two-dimensional filters are replaced by one-dimensional filters. 


GANs trained using the value function proposed in the original GAN paper \cite{GAN} are often unstable, and mode collapse can occur when the generator architecture is varied. Therefore, we use a stable cost function introduced in WGAN \cite{wasserstein}. In this cost function, we minimize the Wasserstein-1 distance between data distribution $p_{data}$(x) and model distribution (Equation \ref{eq:2}). Model distribution is implicit in the second part of the equation because $G$(z) represents the mapping from a latent vector $z$ with distribution $p_z$(z) to the data space. In WGAN, the discriminator network $D_{WGAN}$ gives a score based on the realness of the given image instead of predicting the probability that $x$ comes from the real distribution. In Equation \ref{eq:2}, $E$ represents expectation.
{
\begin{equation}\label{eq:2}
\begin{aligned}[b]
    V_{WGAN}(D_{WGAN},G)=E_{x \sim p_{data}{(x)}}[\log{D_{WGAN}(x)]} \\
    - E_{z \sim p_{z}{(z)}}{[\log{D_{WGAN}(G(z))}].}
\end{aligned}
\end{equation}
}

\begin{figure}[t] 
	\centering
	\includegraphics[width=2in]{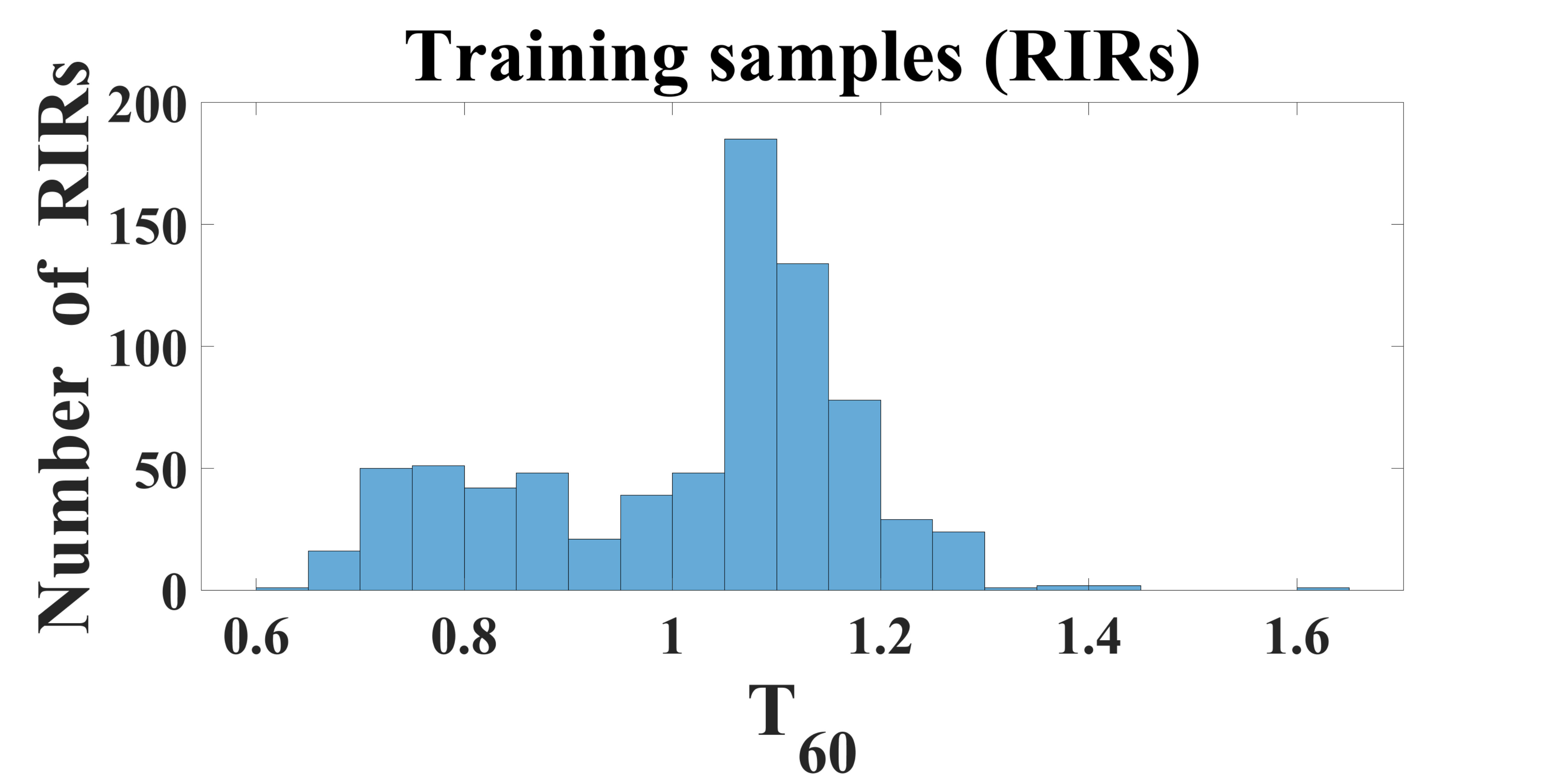}
	\includegraphics[width=2in]{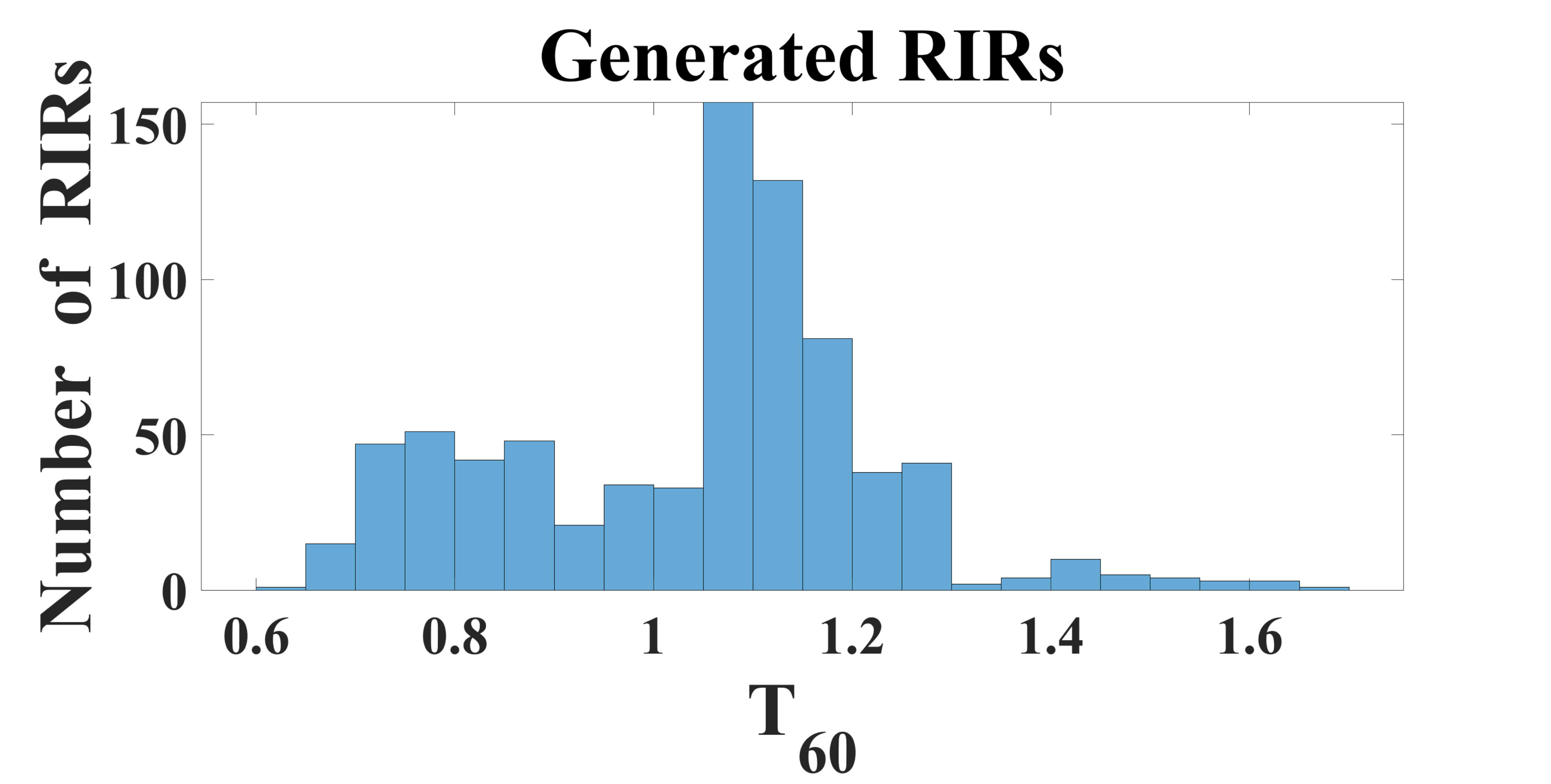}	
	\caption{$T_{60}$ distribution of training samples and $T_{60}$ distribution of RIRs generated using our IR-GAN with the constraint.}
	\label{fig_3}
\end{figure}

\subsection{Constrained RIR Generation}
\label{ssec:IR_Generation}

In our approach, we train a GAN to learn the mapping from the 100-dimensional latent vector $z$  drawn from a Gaussian distribution to the RIR in data space. As the number of real-world RIR datasets is limited, we propose a constrained generation of RIRs from the generator network.

There is an infinite possibility to generate a 100-dimensional vector where each dimension can take any floating-point number between -1 and 1. Since we train GAN with a limited number of RIRs in real RIR datasets (BUT ReverbDB \cite{ButReverb} contains less than 2000 RIRS.), there is a chance that some of the latent vectors map to noisy RIRs. For example, GAN may generate RIRs with unrealistically large $T_{60}$ values. In Figure \ref{fig_2}, we can see a noisy RIR generated without any constraint. The generated noisy RIR with a $T_{60}$ value of around 3 has many horizontal artifacts around 700ms. 

To prevent such mappings, we calculate the key acoustic parameters of the training samples ($T_{60}$, DRR, CTE, and EDT) and use them to generate histograms. Later, we generate RIRs by constraining them to fit the distribution of key acoustic parameters as the training samples. In this way, we can avoid noisy mapping to a greater extent. Since it is difficult to match the exact distributions of the training samples, we relax GAN to generate samples closer to the distribution with low probability. Figure \ref{fig_3} depicts the $T_{60}$ distribution of the training samples and the $T_{60}$ distribution of RIRs generated with the constraint.


\section{Experiments and Results}
\label{sec:experiment}

We evaluate the effectiveness of our proposed approach by conducting far-field automatic speech recognition (ASR) experiments using the modified Kaldi LibriSpeech ASR recipe\footnote{\label{myfootnote}https://github.com/RoyJames/kaldi.}. We use augmented far-field speech to train and test the Kaldi LibriSpeech ASR model and evaluate the benefits in the following manner. First, we compare the performance of our proposed IR-GAN with the state-of-the-art synthetic RIR generator \cite{Diffuse_Acoustic_Simulation}. Second, we evaluate the robustness of our IR-GAN when we train the GAN on one dataset \cite{ButReverb} and test the GAN on another dataset \cite{AIR} from different acoustic environments. We use word error rate to evaluate the performance.

\begin{table}[t]
    \setlength{\tabcolsep}{2pt}
	\caption{Different RIRs used in our experiment.}
	\label{table_symbol}
	\centering
	\begin{tabular}{@{}llr@{}}	
		\toprule
		RIR & Description \\
		\midrule
		BUT & Real-world RIRS from the BUT ReverbDB dataset \cite{ButReverb}.\\
		AIR & Real-world RIRS from the AIR \cite{AIR} dataset.\\
		GAS & Simulated RIRs using the acoustic simulator \cite{Diffuse_Acoustic_Simulation}.\\
	    GAN.C & RIRs generated using our IR-GAN with constraint (\S~\ref{ssec:IR_Generation}).\\
	    GAN.U & RIRs generated using our IR-GAN without any constraint.\\
		\bottomrule
	\end{tabular}
\end{table}

\begin{table}[t]
    \setlength{\tabcolsep}{2pt}
	\caption{Detailed information about the augmented dataset generated using different RIRs (Table \ref{table_symbol}). GAN.C+GAS indicates an equal mixture of GAS and GAN.C synthesized RIRs. 2*GAN.C contains twice the number of RIRs when compared to GAN.C.}
	\label{table1}
	\centering
	\begin{tabular}{@{}llllr@{}}	
		\toprule
		Dataset & RIR & Hours & \#RIRs  & LibriSpeech Dataset\\
		\midrule
        \textbf{Test} & BUT & 5.4  & 242 & test-clean\\
        \textbf{Dataset} & AIR & 5.4  & 68 & test-clean \\
        \midrule
        & BUT & 460 & 773  & train-clean-\{100,360\}\\
        & GAS & 460  & 773 & train-clean-\{100,360\}\\
		\textbf{Training}& GAN.C & 460  & 773  & train-clean-\{100,360\}\\
		\textbf{Dataset} & GAN.U & 460  & 773  & train-clean-\{100,360\}\\
		& GAN.C+GAS & 460 & 1546   & train-clean-\{100,360\}\\
		& 2*GAN.C & 460 & 1546   & train-clean-\{100,360\}\\
		\bottomrule
	\end{tabular}
\end{table}


\subsection{Data Preparation}
\label{ssec:data}
As proposed in \cite{low-frequency_zhenyu}, we generate far-field speech from clean LibriSpeech by convolving it with RIRs and adding environmental noise. Since we mainly focus on the quality of the synthesized RIRs, we use the same environmental noise from the BUT ReverbDB \cite{ButReverb} for training and test set generation.

We use real RIRs from the BUT ReverbDB dataset \cite{ButReverb} and the AIR \cite{AIR} dataset to conduct our experiments. BUT ReverbDB consists of 1891 RIRs and 9114 environmental noises covering nine different rooms. 
To make fair comparisons, we use the 1209 BUT ReverbDB RIRs picked in \cite{low-frequency_zhenyu}. The AIR dataset consists of 344 real RIRs from 6 different rooms. From these, we select 68 RIRs from the 4 rooms (studio booth, office room, lecture room, and meeting room) mentioned in \cite{AIR}. We split 1209 real-world RIRs from the BUT ReverbDB dataset into subsets of \{773,194,242\} to generate training, development, and test far-field speech datasets. To test the robustness of our proposed approach, we use 68 RIRs from the AIR dataset.

Table \ref{table_symbol} describes different RIRs used to create far-field speech datasets in our experiment. Table \ref{table1} shows the detailed composition of the augmented datasets. The augmentation process does not change the overall duration of the original LibriSpeech dataset.



\subsection{Synthetic RIR generation}
\label{ssec:exp_IR_generate}
For a fair comparison, we use the same synthetic RIRs generated using the state-of-the-art geometric acoustic simulator \cite{Diffuse_Acoustic_Simulation} as the previous benchmark. The geometric acoustic simulator synthesizes RIR using the meta-info provided in BUT ReverbDB. The meta-info includes room dimensions and loudspeaker and microphone locations. Therefore, the synthesized RIRs for the training and development sets mimic real RIR training and development sets to some extent.

We train our IR-GAN with 967 real RIRs from the BUT ReverbDB dataset. They are composed of real-world RIRs allocated for training and development. We generate synthetic RIRs with and without the constrained RIR generation process in \S~\ref{ssec:IR_Generation}.

\subsection{ASR Experiment}
\label{ssec:ASR}
We use the modified Kaldi LibriSpeech ASR recipe to conduct our ASR experiments. We train time-delay neural networks \cite{TDNN} for each of our augmented far-field training sets. We extract the identity vectors \cite{ivector} (i-vectors) of the real-world far-field test set and decode them using large four-gram (fglarge), large tri-gram (tglarge), medium tri-gram (tgmed), and small tri-gram (tgsmall) phone language models. We also do online decoding on the tgsmall phone language model. In online decoding, extracted features are passed in real-time instead of waiting until the entire audio is captured. Word error rate (WER) of each of the language model is used to evaluate the synthesized RIRs.

All the training and testing is done on 32 Intel(R) Xeon(R) Silver 4208 CPUs @ 2.10 GHz and 2 GeForce RTX 2080 Ti GPUs. For a fair comparison, we generate all the results from the same environment. It takes around four days to prepare the dataset and conduct each experiment on the Kaldi toolkit.



\begin{table}[t]
   \setlength{\tabcolsep}{2pt}
	\caption{Far-field automatic speech recognition results obtained from the far-field LibriSpeech test set. In this table, *BUT and *AIR represent far-field test sets generated using real RIRs from the BUT ReverbDB and AIR datasets, respectively. clean* represents clean speech. WER is reported for the tri-gram phone (tglarge, tgmed, tgsmall) and four-gram phone (fglarge) language models, and online decoding using tgsmall. Best results in each comparison are marked in \textbf{bold}.}
	\label{table2}
	\centering
	\begin{tabular}{@{}lllllr@{}}	
		\toprule
		\multirow{2}{34mm}{Experiment Setup\\ (training set)\at(test set)}
			& \multicolumn{5}{c}{Test Word Error Rate (WER) [\%]}\\
\cmidrule(r{4pt}){2-6} 
			& fglarge & tglarge & tgmed & tgsmall & online\\
		\midrule
		clean\at BUT (Baseline) & 77.15 & 77.37 & 78.00 & 78.94 & 79.00\\
		BUT \at BUT (Oracle) & 12.40 & 13.19 & 15.62 & 16.92 & 16.88\\
		\midrule
		GAS\at BUT \cite{Diffuse_Acoustic_Simulation} & 16.53 & 17.26 & 20.24 & 21.91 & 21.83\\
		GAN.U\at BUT & 19.71 & 20.74 & 24.27 & 25.93 & 25.90\\
		GAN.C\at BUT & \textbf{14.99} & \textbf{15.93} & \textbf{18.81} & \textbf{20.28} & \textbf{20.24} \\
		\midrule
		2*GAN.C\at BUT & 14.86 & 15.69 & 18.50 & 20.25 & 20.17 \\
		GAN.C+GAS\at BUT & \textbf{14.16} & \textbf{14.99} & \textbf{17.56} & \textbf{19.21} & \textbf{19.21}\\
        \midrule
		clean\at AIR& 26.79 &27.40 & 29.64 & 30.88 & 31.15 \\
		GAN.C\at AIR & \textbf{7.71} & \textbf{8.03} & \textbf{9.88} & \textbf{11.11} & \textbf{11.08} \\
		\bottomrule
	\end{tabular}
\end{table}
\subsection{Results}
\label{ssec:results}
Table \ref{table2} presents the ASR test word error rate (WER) for far-field speech generated using the BUT ReverbDB \cite{ButReverb} and AIR \cite{AIR} datasets. WER is calculated for four different phone language models (fglarge, tglarge, tgmed, and tgsmall) in Kaldi as well as for online decoding using a tgsmall phone language model. 

We use WER to measure the robustness of the trained model. A lower WER indicates that the trained model shows superior accuracy in test conditions. Robustness depends on the model architecture and the input data used to train the model. In our experiments, we keep the model constant while we train with different datasets. Different training datasets are created by convolving the same LibriSpeech speech corpus with different RIRs. Therefore, the robustness of the model is only affected by the RIRs being used to generate the training datasets. As expected, a significantly high WER is reported when we train our baseline model on clean speech and test on the real RIRs. The lowest test WER is reported when we train and test on the real RIRs.


In Table \ref{table2}, we can see that the proposed IR-GAN (GAN.C) gives a lower WER than the state-of-the-art geometric acoustic simulator (GAS). Lower WER indicates that the RIRs synthesized using IR-GAN are more realistically synthesized than the RIRs computed using the physically-based acoustic simulator. When we look at the WER for the fglarge model, we can see that our proposed IR-GAN gives an 8.95\% lower error rate than the GAS. We can see that the RIRs generated using the unconstrained IR-GAN (GAN.U) performs poorly in our far-field speech recognition experiment. Therefore our constrained RIR generation approach is important to eliminate noisy RIR generation.

{\bf Hybrid Combination:} Because the IR-GAN and GAS try to mimic real RIRs using two different approaches, we evaluate the WER when trained using a combination of synthetic RIRs generated using the IR-GAN and GAS. We observe that there is a further drop of up to 5\% in WER compared to doubling the synthetic RIRs from the IR-GAN. The drop in WER indicates that we can boost the robustness of ASR systems by combining RIRs generated from our IR-GAN and GAS.

In practical scenarios, we do not have real RIRs from the acoustic environment where we need the capabilities for far-field ASR. Therefore, we augment RIRs using our IR-GAN trained with real RIRs from BUT ReverbDB \cite{ButReverb}. Then we train the Kaldi LibriSpeech ASR model on far-field speech generated using the augmented RIRs and test this ASR model on the far-field speech augmented using the AIR dataset \cite{AIR}.
We can observe around a 19\% absolute reduction in error when compared to training an ASR model with clean speech. 

\section{Discussion and Future Work}
\label{ssec:discussion}

In this paper, we present an IR-GAN to generate realistic RIRs. Our proposed approach outperforms the state-of-the-art geometric acoustic simulator (GAS) by up to 8.95\% in far-field ASR tests. When we combine our RIRs with RIRs generated using GAS, we can see a total reduction in word error rate by up to 14.3\% in far-field ASR tests. This reduction in word error rate indicates that synthetic data generated using IR-GAN and GAS can be combined to boost the performance of far-field ASR systems. We have tested our approach only on indoor scenes, and extending it to outdoor scenes is a good topic for future work.


\bibliographystyle{IEEEtran}

\bibliography{mybib}


\end{document}